\documentclass[aps,pra,epsfigure,twocolumn,showpacs,superscriptaddress]{revtex4}
\usepackage{color}

\usepackage{amsmath}
\usepackage{amssymb,amsthm}
\usepackage{verbatim}
\usepackage{epsfig}
\usepackage{epstopdf}

\begin{document}

\title{The hierarchies of ``witnesses" and the properties and characterization of entangled states as super entanglement witnesses}

\author{Bang-Hai~Wang
\footnote{Correspondence to:
wangbanghai@gmail.com}}%
\affiliation{School of Computer Science and Technology, Guangdong University of Technology, Guangzhou 510006, People's Republic of China}
\affiliation{Department of Computer Science, and Department of Physics and Astronomy, University College London, Gower Street, WC1E 6BT London, United Kingdom}

\date{\today}

\begin{abstract}

{\bf Quantum entanglement lies at the heart of quantum mechanical and quantum information processing. Following the question who \emph{witnesses} entanglement witnesses, we show entangled states play as the role of super entanglement witnesses. We show separable states play the role of ``super super entanglement witnesses" and ``witness" other observables than entanglement witnesses. We show that there exists a hierarchy structure of witnesses and there exist ``witnesses" everywhere. Furthermore, we show the properties and characterization of entangled states as super entanglement witnesses. By the role of super witnesses of entangled states, we immediately find the question when different entanglement witnesses can detect the same entangled states [{\it Phys. Lett. A }{\bf 356} 402 (2006)] is the same as the question when different entangled states can be detected by the same entanglement witnesses [{\it Phys. Rev. A} {\bf 75} 052333 (2007)]. By the role of ``witnesses", we define finer entangled states and optimal entangled states. The definition gives a nonnumeric measurement of entanglement and an unambiguous discrimination of entangled states, and the procedure of optimization for a general entangled state $\rho$ is just finding the best separable approximation (BSA) to $\rho$ in [ {\it Phys. Rev. Lett.} {\bf 80} 2261 (1998)].}

\end{abstract}


\maketitle

\noindent Quantum correlations, especially quantum entanglement have been recognized as a novel resource that may be used for tasks that are neither impossible or very inefficient in the classical realm \cite{Plenio07}. However, quantum entanglement has not been fully understood. The effective method has not yet been found to detect whether or not a given state is entangled. Even though it is known to be entangled, but the amount of entanglement cannot easily be determined for a general mixed entangled state. A remarkable research effort has been devoted to detecting and quantifying it \cite{Guhne09,Horodecki09}. The method of entanglement witnesses is considered to be currently the most important and the best method to detect entanglement \cite{Augusiak11}.

It is well known that the set of separable states is convex and compact. Following from a consequence of the Hahn-Banach theorem \cite{Edwards65}, there exists at least one operator to detect it for any entangled state \cite{M.Horodecki96}, as shown in Fig. 1 (a). This operator was coined entanglement witnesses by Terhal who stressed their physical importance as entanglement detectors \cite{Terhal00, Horodecki09}.


An entanglement witness is a Hermitian operator, $W=W^\dag$, such that

(i) $\text{tr}(W\sigma)\geq0$ for an arbitrary separable state $\sigma$, and

(ii) there exists an entangled state $\pi$ such that $\text{tr}(W\pi)<0$.

In that case an entanglement witness $W$ detects an entangle state $\pi$, we say that
the entanglement witness $W$ ``witnesses" the quantum state $\pi$.
The entanglement witness has been widely investigated by many people \cite{Lewenstein98,Doherty04,Wu06,Wu07,Sperling09,Chruscinski09,Hou10a,Chruscinski10,Wang11,Chruscinski14}. Unfortunately, to construct the entanglement witness for an entangled state is a difficult task, and the determination of entanglement witnesses for all entangled states is a nondeterministic polynomial-time (NP) hard problem\cite{Gurvits04,Doherty04,Hou10a}.

At the outset, the entanglement witness was introduced because we cannot directly detect entanglement. Now, how to construct entanglement witnesses in general, and finding the minimal set of them that allows detection of all entangled states have become the most challenging open questions \cite{Lewenstein01}. A natural question arises immediately. Who ``witnesses" entanglement witnesses? (see Fig. 1 (b).)

\begin{figure}[htbp]
\epsfig{file=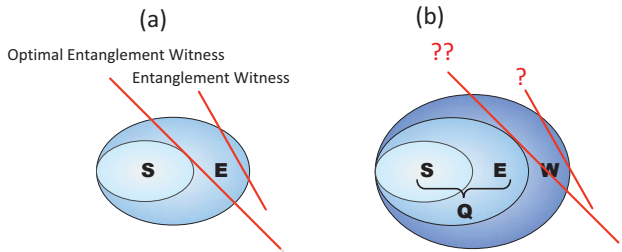,width=.95\columnwidth}
\caption{(Color online) We denote $\mathbf{S}$ the set of all separable states, $\mathbf{E}$ the set of all entangled states, $\mathbf{W}$ the set of all entanglement witnesses, and $\mathbf{Q}=\mathbf{S}\cup\mathbf{E}$ the set of all quantum states. (a) Schematic picture of the set of all states as a convex set. Entanglement witnesses witness entangled states. (b) Schematic picture of the set of all states as a convex subset. Who witnesses entanglement witnesses?}
\label{fig1}
\end{figure}

It is known that the set of quantum states (separable states and entangled states) is also convex and compact. Following from a consequence of the Hahn-Banach theorem (For Gaussian states in systems with continuous variables, the sets will be closed and convex, but not necessarily bounded \cite{Brub2002}), there is at least a ``super witness" witnessing the entanglement witness, as shown in Fig. 1 (b).
For a super witness $\Pi$,

(i') $tr(\Pi\rho)\geq0$ for an arbitrary quantum state $\rho$ (entangled or not);

(ii') there is at least an entanglement witness $W$, $tr(\Pi W)<0$.


\vspace{8pt}

Interestingly, the one who satisfies the two conditions and plays the role of ``super witness" is not other but the entangled state! Entanglement witnesses ``witness" entangled states, and entangled states and ``witness" entanglement witnesses!
As shown in Fig. 2 (a).

\begin{figure}[htbp]
\epsfig{file=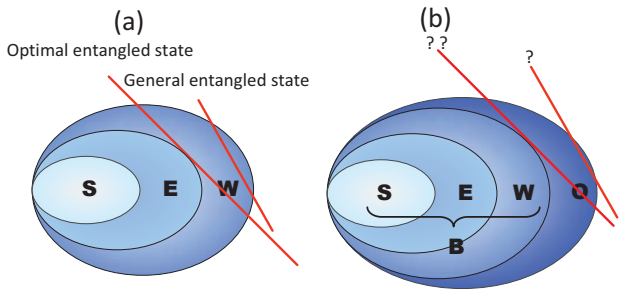,width=.95\columnwidth}
\caption{(Color online) We denote $\mathbf{B}=\mathbf{S}\cup\mathbf{E}\cup\mathbf{W}$ the set of all block-positive operators. (a) Schematic picture of the set of all states as a convex subset. Entangled states witness entanglement witnesses. (b) Schematic picture of the set of all block-positive operators as a convex subset. Who witnesses other observables than entanglement witnesses?}
\label{fig2}
\end{figure}

This answer is not difficult to get, but it is unexpected. The role of super entanglement witnesses motivates the study of entangled states as a new way to study the properties of entangled states. The entangled states own the same or the dual characterization and properties as entanglement witnesses. 
It seems that the super-witness properties and characterization of entangled states have been ignored for a long time. For simplicity, in the following we shall freely use the super witnesses and the entangled states.

Furthermore, we can also ask whether or not there exist ``super super" entanglement witnesses, as shown in Fig. 2 (b). To investigate these super witnesses, let us consider the set of bounded Hermitian operators, which have
positive expectation values for separable states
\begin{equation}
\mathbf{\mathbf{B}}=\{b\,|\, b=b^\dag, |tr(\sigma b)\geq0\}
\end{equation}
where $\sigma$ is any separable state. The set $\mathbf{B}$ is called the set of block-positive \cite{Chruscinski14} or partial positive
operators \cite{Wu06,Sperling09}. In standard quantum mechanics, all observables are mathematically denoted by Hermitian operators. We can also separate other observables from entanglement witnesses.

We can easily conclude that the set of block operators $\mathbf{B}$ is also convex and compact. One may still ask who is this super super entanglement witness that separates other observable from the quantum state (separable and entangled) and entanglement witnesses, as shown in Fig. 2 (b).

To investigate these super super witnesses, let us consider the set of bounded Hermitian operators
\begin{equation}
\mathbf{H}=\{h|h=h^\dag\}, \mathbf{O}=\mathbf{H}-\mathbf{S}-\mathbf{E}-\mathbf{W}.
\end{equation}

For a super super witness $\Upsilon$,

(i'') $tr(\Upsilon b)\geq0$ for any block operator $b\in \mathbf{B}$ (a quantum state or an entanglement witness);

(ii'') there is at least a given non-entanglement-witness observable $o\in\mathbf{O}$ such that $tr(\Upsilon o)<0$.

\begin{figure}[htbp]
\epsfig{file=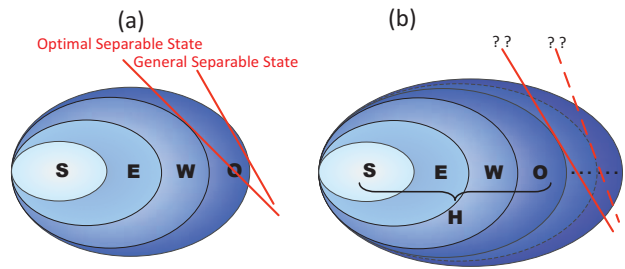,width=.95\columnwidth}
\caption{(Color online) We denote $\mathbf{H}$ the set of all Hermitian operators. (a) Schematic picture of the set of all block-positive operators as a convex subset. Separable states witness other observables than entanglement witnesses. (b) Schematic picture of the set of all Hermitian operators as a convex subset. Who witnesses non-hermitian operators? Are there infinite ``super" witnesses, physically, or mathematically? }
\label{fig3}
\end{figure}

We can easily conclude that the super super entanglement witness is just separable states, as shown in Fig. 3 (a). In other words, separable states separate entanglement witnesses from other observables than entanglement witnesses. Sometimes, some non-Hermitian operators \cite{Moiseyev11} are also required to measure. We can find the witnesses of some non-Hermitian operators, as shown in Fig. 3 (b).

Mathematically, we can construct more and more convex and compact set
such that it includes the set of Hermitian operators. There exists more and more convex and compact sets, and there exists a hierarchy structure of witnesses, as shown in Fig. 3 (b).

If an entanglement witness can be written in the form
\begin{equation}\label{DecomposableEW}
W_d=aP+(1-a)Q^\Gamma,
\end{equation}
where $a\in[0,1]$, $P\geq0$, and $Q\geq0$, the entanglement witness is called decomposable \cite{Lewenstein00}. If it doesn't admit this form, it is called indecomposable. Clearly, the set of decomposable entanglement witnesses $\mathbf{W_d}$ is convex and compact \cite{Lewenstein00}. By Hahn-Banach theorem, there exists at least an operator $\rho$ such that (i) $tr[\rho W_d]\geq0$ for all decomposable entanglement witnesses $W_d$; and (ii) $tr(\rho W_{nd})<0$ for an indecomposable entanglement witness $W_{nd}$. In other words, there exist ``witnesses" separate indecomposable entanglement witnesses from decomposable entanglement witnesses. It is easy to know that these ``witnesses" are just bound entangled states. Furthermore, arbitrary quantum state can be written as the form of Eq. (\ref{DecomposableEW}), the set $\mathbf{Q}\cup \mathbf{W_d}$ is also convex and compact \cite{Sarbicki07}.





In fact, there exist many many ``witnesses" in nature, even between any one state and another different state because we can consider one state as a closed, convex set with only one element. ``Witnesses" are everywhere!

\vspace{4pt}

Here we show the properties and characterization of entangled states as super entanglement witnesses. Let us first focus on the known question when different entanglement witnesses can detect the same entangled states.

\textbf{Lemma 1.} \cite{Wu06} (see also\cite{Hou10a}) There exists $\rho$ detected by $W_1$ and $W_2$ if and only if for any $\lambda\in [0,1]$, $W=\lambda W_1+(1-\lambda)W_2$ is not a positive operator (in other words, $W=\lambda W_1+(1-\lambda)W_2$ must be an entanglement witness because $\text{tr}(W_1\sigma)\geq0$, $\text{tr}(W_2\sigma)\geq0$, and $\text{tr}(W\sigma)\geq0$ for an arbitrary separable state $\sigma$).

Since entangled states are also (super entanglement) witnesses, this question can change into when different super entanglement witnesses (entangled states) can detect the same entanglement witness. By the super-witness property of entangled states, we can immediately get the following result.

\textbf{Corollary 1.} There exists entanglement witness $W$ detected by a super entanglement witness (entangled state) $\Pi_1$ and a super entanglement witness (entangled state) $\Pi_2$ if and only if for any $\lambda\in [0,1]$, $\Pi=\lambda \Pi_1+(1-\lambda)\Pi_2$ is a super entanglement witness (an entangled state).

This recovers the main result of Ref. \cite{Wu07}. Note that the proof procedure of Lemma 1 used only the ``general" properties and characterization of ``witnesses" (see Ref. \cite{Hou10a}, Ref. \cite{Sarbicki07}, and Supplemental Materials).


According to the definition in Ref. \cite{Lewenstein00}, (i)
given an entanglement witness $W$, define $D_W=\{\pi\geq0$, such that $\text{tr}(W\pi)<0\}$, i.e.,
the set of density matrices detected by $W$; (ii) given two entanglement witnesses,
$W_1$ and $W_2$, $W_2$ is finer than $W_1$ if $D_{W_1}\subseteq
D_{W_2}$; and (iii) $W$ is an optimal entanglement witness (OEW) if
there exists no other finer witness. This definition shows the ``detection power" of entanglement witnesses.

Following the definition above, we define the following. Given a super entanglement witness $\rho$, define $S_\rho=\{W\ngeq0,\mbox{ such that } \langle W\rangle_\rho<0\}$; that is the set of operator witnessed by $\rho$. Given two super witnesses, $\rho_1$ and $\rho_2$, we say that $\rho_2$ is finer than $\rho_1$, if $S_{\rho_1}\subseteq S_{\rho_2}$; that is, if all the operators witnessed by $\rho_1$ are also witnessed by $\rho_2$. We say that $\rho$ is an optimal super witness if there exists no other super witness which is finer.

Once we replace ``finer" with ``more entangled", we can immediately get an entanglement measure in theory. Given an entangled $\rho$, $S_\rho=\{W\ngeq0,\mbox{ such that } \langle W\rangle_\rho<0\}$; that is the set of operator witnessed by $\rho$. Given two entangled states, $\rho_1$ and $\rho_2$, we say that $\rho_2$ is more entangled than $\rho_1$, if $S_{\rho_1}\subseteq S_{\rho_2}$; that is, if all the operators witnessed by $\rho_1$ are also witnessed by $\rho_2$. We say that $\rho$ is an optimal entangled if there exists on other super witness which is more entangled.

This definition shows the ``witnessing power" of entangled states. Using an entanglement witness, we can quantify the entanglement content via
\begin{equation}
\label{Fernando}
E(\rho)=\max\{0,-\min_{W\in M} \text{tr} (W\rho) \},
\end{equation}
where $M$ is the intersection of the set of entanglement witnesses with some other set $C$ such that $M$ is compact, (see~\cite{Brandao05}). However, our entanglement measure quantifies by the number of witnesses who are witnessed by the entangled states other than the numeric value.

Generally, there are two categories of measure entanglement \cite{Horodecki09}. One is based on definitions of operational tasks. This entanglement measure owns direct physical implication, such as entanglement cost, entanglement of distillation, but it is difficult to compute. Another one is based on the view of axiomatic \cite{Vedral97}. The entanglement measure is often operator functions satisfying several basic properties of operator functions, such as concurrence, negativity, entanglement of formation, and so on. It is also called entanglement monotone.

Almost all known criteria, however, map an entangled state to a real number (sometimes between 0 and 1 for comparison). Generally, it will depend on what criterion is adopting that if a state is more entangled than another. Therefore, there exists the possibility that two different entangled states have the same value. There also exists the possibility that a criterion indicates $\rho_1$ is more entangled than $\rho_2$ but another criterion shows $\rho_2$ is more entangled than $\rho_1$.
We can quantify entanglement for an entangled state with entanglement witnesses, who is ``super" witnessed by the entangled state. For two different entangled states, there exist different sets of witnesses to character entangled states, respectively.

We argued that if there exists no ``finer" relation between two super witnesses (entangled states), the two entangled states cannot be simply told which one is more entangled than the other, just as we cannot simply tell who is ``finer" between the badminton world champion and the tennis world champion.

Consider the Werner state~\cite{Werner89}
\begin{equation}
\label{werner}
\pi_p=p|\psi\rangle\langle\psi|+(1-p)I/4,
\end{equation}
where $|\psi\rangle=\frac{1}{\sqrt{2}}(|00\rangle+|11\rangle)$ and $0\leq p\leq1$. It is well known that $\pi_p$ is entangled for $\frac{1}{3}<p\leq1$. Since the entanglement witness $W=|\varphi\rangle\langle\varphi|^\Gamma$ is optimal, the eigenvector of the negative eigenvalue is just $|\psi\rangle$, where $|\varphi\rangle=\frac{1}{\sqrt{2}}(|10\rangle-|01\rangle)$.

 We can easily determine
 \begin{equation}
\mathbf{S}_{\pi_p}=\{W=q|\varphi\rangle\langle\varphi|^\Gamma+(1-q)\rho|tr(\pi_pW)<0 \},
\end{equation}
where $0\leq q\leq1$, $\rho$ is a quantum state and $\pi_p$ is super entanglement witness for $\frac{1}{3}<p\leq1$, $\pi_s$ is finer (more entangled) than $\pi_t$ for $\frac{1}{3}<s<t\leq1$ and $|\psi\rangle$ is an optimal entangled state.

Since $S_{\rho_1}=S_{\rho_2}$ if and only if $\rho_1=\rho_2$ (see Supplemental Materials), these sets not only can quantify entanglement but also can distinguish different entangled states, and we can perfectly distinguished entangled states by the sets of entanglement witnesses in theory.

\textbf{Lemma 2.} Entangled states are perfectly distinguishable by the sets of entanglement witnesses super-witnessed by entangled states.

We can also define a separable state is finer than another one and the optimal separable state. Similarly, we have the following result.

\textbf{Lemma 3.} Separable states are perfectly distinguishable by the sets of the observables other than entanglement witnesses super-super-witnessed by separable states.

We summarize Lemma 2 and Lemma 3, and we have a perfectly distinguishable of quantum states.

\textbf{Theorem 1.} Quantum states are perfectly distinguishable by the sets of the observables super-witnessed or super-super-witnessed by quantum states.

If one subtracts a projector onto a product vector from a positive partial transposition entangled state (PPTES), the resulting operator is no longer a PPTES, the PPTES is called an edge state \cite{Lewenstein98,Lewenstein01,Lewenstein00,Sanpera98}, because it lies in the edge between PPTES's and entangled states with non-positive partial transposition. Although the range criterion immediately implies that an edge operator is necessarily entangled \cite{Chruscinski14}, our result shows edge states are optimal entangled. In other words, edge states constitute a part of optimal entangled states. We can easily conclude the following results.


\textbf{Theorem 2.} If a quantum state whose range does not contain any product vector $|e,f\rangle$, it is optimal entangled state.

Therefore, we can derive a direct criterion, which is different from the ones of optimal entanglement witnesses, to determine whether or not an entangled state is optimal.

\textbf{Theorem 3.} If the support of an entangled state $\pi$ does not contain any product vector or $Support(\pi)$ is a completely entangled subspace (CES) \cite{Parthasarathy04}, which does not contain any product state, then $\pi$ is optimal.

\textbf{Corollary 2.} All mixed states on a CES are optimal entangled states.

According to the result, we can construct optimal entangled states by mixing the entangled basis in CES. Since the maximum dimension of subspace in $\mathcal{H}=\mathcal{H}_2\otimes\mathcal{H}_2\otimes\cdots\mathcal{H}_k$ is $d_1d_2\cdots d_k-(d_1+d_2+\cdots\cdots+d_k)+k-1$ \cite{Parthasarathy04}, we can immediately conclude the following result in $\mathcal{C}^2\otimes\mathcal{C}^2$.

\textbf{ Corollary 3:} A super entanglement witness in $\mathcal{C}^2\otimes\mathcal{C}^2$ is optimal if and only if it is a single pure entangled state.

It recovers one of the main result in Ref. \cite{Lewenstein98}.

Since an unextendible product basis (UPB) for a multipartite quantum system is an incomplete orthogonal product basis whose complementary subspace contains no product state, we can construct optimal entangled states by the complementary subspace of UPB.

\textbf{Corollary 4.} The state that corresponds to the uniform mixture on the space complementary to a UPB $\{\psi_i: i=1, \ldots, n\}$ in a Hilbert space of total dimension D
\begin{equation}
\bar{\pi}=\frac{1}{D-n}(1-\sum_{j=1}^n|\psi_j\rangle\langle\psi_j|),
\end{equation}
is an optimal entangled state.

By the result in \cite{Bennett99}, we can know it is an optimal bound entangled state.


\vspace{8pt}

\noindent

According to results on entanglement witnesses, we can also restrict ourselves to the study of optimal entangled states when we investigate into entangled states. For that we need conditions that an entangled state is finer than another one, and criteria to determine when an entangled state is optimal. Fortunately, these criteria and conditions are similar to these of entanglement witnesses \cite{Lewenstein00}. For complete, we attached them in the Supplemental Materials, we can also see Ref. \cite{Sarbicki07}.

These results tell us, based on the technique of ``subtracting projectors on product vectors", we can optimize an entangled state by subtracting separable states from the entangled state. Once we subtract
any separable positive operator from it, the resulting operator is not positive or positive but not entangled, the entangled state is optimal. Interestingly, the procedure of optimization for a general entangled state $\rho$ is merely to find the BSA (best separable approximation) to $\rho$ \cite{Lewenstein98,Sanpera98}. This procedure, however, is not very practical. Lewenstein et al. showed how to circumvent the drawbacks in practice in Ref. \cite{Lewenstein00}. Every entangled state which is not optimal can be optimized.
 Note that this technique not only can optimize a non-optimal entangled state but also optimize a ``mixed" separable state to
an optimal entangled state. It is not difficult to conclude that the optimization of quantum states to optimal separable states will be the procedure of subtracting optimal entangled states.

Once we find there exist ``witnesses" between two sets, how to generally construct these ``witnesses"? If $S_1$, $S_2$ are convex closed sets in a real Banach space and one of them is compact, there exists a continuous functional $f$ and $c\in \mathbb{C}$ such that for all pairs $e_1\in S_1$, $e_2\in S_2$, we have $f(e_1)<c\leq f(e_2)$ \cite{M.Horodecki96}. It is known that any continuous functional $f$ on a Hilbert space can be represented by a vector from this space. For any linear functional $g$, acting on trace class operators $\rho$, can be written as $g(\rho)=tr(\rho H)$ for a bounded, Hermitian operator $H$ \cite{Bruns16}. Therefore, we find Hermitian operators, who ``witnesses" $\rho_2\in S_2$, such that $min\{tr(e_2 H)\}<tr(\rho_2 H)\leq min\{tr(e_1 H)\}$. Moreover, the optimization problem under certain constraints can be solved by the method of Lagrange's multipliers (see \cite{Sperling09,Bruns16}).





Like entanglement witnesses, we can also introduce the notions of extremal, exposed entangled states and so on.
Some methods to construct entanglement witnesses may be used for construction of natural entangled states. Under the Jamio{\l}kowski-Choi isomorphism, positive but not completely positive maps correspond to entanglement witnesses (matrices). we can investigate the properties and characterization of entanglement by the completely positive maps, which correspond to entangled states, separated from the completely positive maps.
In all, all properties and characterization of entanglement witnesses should be investigated to determine whether or not they can be employed to entangled states. Some of them \cite{Sarbicki07} can be directly employed to entangled states, some of them may be analyzed with the intrinsic order of entanglement. A lot of interesting questions left deserves further study.




{\bf Acknowledgements}

We are grateful to Christopher Perry, Dan Brown, Jonathan Oppenheim, Fernando Brand\~{a}o, Simone Severini, Dong-Yang Long, and Guang Ping He for very helpful discussion and suggestions. We thank the Department of Computer Science and the Department of Physics \& Astronomy
at University College London for their
hospitality, where part of this work was carried out while Wang was an academic visitor. We thank the Isaac Newton Institute for Mathematical
Sciences, Cambridge for their hospitality during the programme ``Mathematical
Challenges in Quantum Information¡±, where part of this work was carried out. This work is supported by the National Natural
Science Foundation of China under Grant No. 61272013 and was supported by the National Natural
Science Foundation of Guangdong province of China under Grant No. s2012040007302.

{\bf Supplementary Information}

\section{optimal entangled states}

\textbf{ Lemma A1:} Let $\rho_2$ be finer than $\rho_1$ and
\begin{equation}
\label{delta}
\delta\equiv \inf_{W_1\in D_{\rho_1}} \left|
\frac{\langle{W_1}\rangle_{\rho_2}}{\langle{W_1}\rangle_{\rho_1}} \right|.
\end{equation}
Then we have the following:

\begin{description}
\item [(i)] If $\langle{W}\rangle_{\rho_1}=0$ then $\langle{W}\rangle_{\rho_2}\leq 0$.

\item  [(ii)] If $\langle{W}\rangle_{\rho_1}<0$, then $\langle{W}\rangle_{\rho_2} \leq
\langle{W}\rangle_{\rho_1}$.

\item [(iii)] If $\langle{W}\rangle_{\rho_1}>0$ then $\delta\langle{W}\rangle_{\rho_1}\geq
\langle{W}\rangle_{\rho_2}$.

\item [(iv)] $\delta\geq 1$. In particular,
$\delta=1$ iff $\rho_1=\rho_2$.

\end{description}

{\em Proof:} Since $\rho_2$ is finer than $\rho_1$ we will use the
fact that for all $W\ngeq 0$ such that $\langle{W}\rangle_{\rho_1}<0$ then
$\langle{W}\rangle_{\rho_2}<0$.

{\em (i)} Let us assume that $\langle{W}\rangle_{\rho_2} > 0$. Then we take
any $W_1\in S_{\rho_1}$ so that for all $x\ge 0$, $0\le
\tilde W(x)\equiv W_1+xW \in S_{\rho_2}$. But for
sufficiently large $x$ we have that
$\langle{\tilde W(x)}\rangle_{\rho_2}$
is positive, which cannot be since then $\tilde W(x)
\not{\in} S_{\rho_2}$.

{\em (ii)} We define $\tilde W=W+|\langle{W}\rangle_{\rho_1}|I\ge
0$, where $I$ is the identity
matrix. We have that $\langle{\tilde W}\rangle_{\rho_1}=0$. Using (i) we have
that $0\ge
\langle{W}\rangle_{\rho_2}+|\langle{W}\rangle_{\rho_1}|$.

{\em (iii)} We take $W_1\in S_{\rho_1}$ and define $\tilde W=
\langle{W}\rangle_{\rho_1} W_1 + |\langle{W_1}\rangle_{\rho_1}| W \ge 0$, so that $\langle{\tilde
W}\rangle_{\rho_1}=0$. Using (i) we have $|\langle{W_1}\rangle_{\rho_1}|\langle{W}\rangle_{\rho_2}\le
|\langle{W_1}\rangle_{\rho_2}|\langle{W}\rangle_{\rho_1}$. Dividing both sides by
$|\langle{W_1}_{\rho_1}|>0$ and $\langle{W}\rangle_{\rho_1}>0$ we obtain
\begin{equation}
\frac{\langle{W}\rangle_{\rho_2}}{\langle{W}\rangle_{\rho_1}} \le  \left|
\frac{\langle{W_1}\rangle_{\rho_2}}{\langle{W_1}\rangle_{\rho_1}} \right|.
\end{equation}
Taking the infimum with respect to $W_1\in S_{\rho_1}$ on the
right hand side of this equation we obtain the desired result.

{\em (iv)} From (ii) immediately follows that $\delta\ge 1$. On
the other hand, we just have to prove that if $\delta=1$ then
$\rho_1=\rho_2$ (the only if part is trivial). If $\delta=1$, using
(i) and (iii) we have that
$\langle{W_v}\rangle_{\rho_1}\ge\langle{W_v}\rangle_{\rho_2}$ for all
$W_v=|e,f\rangle\langle e,f|$ projector on a product vector.
Since $\mbox{tr}(\rho_1)=\mbox{tr}(\rho_2)$ we must have $\mbox{tr}[(\rho_1-\rho_2)W_v]=0$
for all $W_v$, since we can always find a product basis in
which we can take the trace. But now, for any given $W\ngeq 0$ we
can define $\tilde
W(x)=W+x I$ such that for large enough $x$, $\tilde W(x)$ is
separable \cite{Wang13}. In that case we have $\langle{\tilde
W(x)}\rangle_{\rho_1}=\langle{\tilde W(x)}\rangle_{\rho_2}$ which implies that
$\langle{\rho}\rangle_{\rho_1}=\langle{\rho}\rangle_{\rho_2}$, i.e. $\rho_1=\rho_2$. $\Box$

\textbf{ Corollary A1:} $S_{\rho_1}=S_{\rho_2}$ if and only if $\rho_1=\rho_2$.

{\it Proof:} We just have to prove the only if part. For that,
we define $\delta$ as in (\ref{delta}). On the other hand,
defining
\begin{equation}
\tilde\delta\equiv \inf_{W_2\in S_{\rho_2}} \left|
\frac{\langle{W_2}\rangle_{\rho_1}}{\langle{W_2}\rangle_{\rho_2}}\right|
\end{equation}
we have that $\tilde\delta\ge 1$ since $\rho_1$ is finer than
$\rho_2$ (Lemma A1(iv)). Equivalently,
\begin{equation}
1\ge \sup_{W_1\in S_{\rho_1}} \left|
\frac{\langle{W_1}\rangle_{\rho_2}}{\langle{W_1}\rangle_{\rho_1}}\right| \ge \delta\ge 1,
\end{equation}
where for the last inequality we have used that $\rho_2$ is finer
than $\rho_1$. Now, since $\delta=1$ we have that $\rho_1=\rho_2$
according to Lemma A1(iv). $\Box$

Next, we introduce one of the basic results of optimal entangled states. It basically
tell us that an entangled state is finer than another one if they differ by projectors on product vectors. That is, if we have an entangled state and we want to find another one
which is finer, we have to subtract a projector on product vectors.

\textbf{ Lemma A2:} $\rho_2$ is finer than $\rho_1$ if and only if there exists
$1>\epsilon\geq 0$ such that $\rho_1=(1-\epsilon)\rho_2+\epsilon P$, where
$P$ is not finer than $\rho_1$ or it is separable.

{\it Proof:} (If) For all $W\in S_{\rho_1}$ we have that
$0>\langle{W}\rangle_{\rho_1}= (1-\epsilon)\langle{W}\rangle_{\rho_2}+\epsilon \langle{W}\rangle_{P}$
which implies $\langle{W}\rangle_{\rho_2}<0$ and therefore $W\in S_{\rho_2}$. (Only
if) We define $\delta$ as in (\ref{delta}). Using Lemma A1(iv) we have
$\delta\ge 1$. First, if $\delta=1$ then according to Lemma A1(iv) we
have $\rho_1=\rho_2$ (i.e., $\epsilon=0$). For $\delta> 1$, we define
$P=(\delta-1)^{-1}( \delta \rho_1-\rho_2)$ and $\epsilon=1-1/\delta>0$.
We have that $\rho_1=(1-\epsilon)\rho_2+\epsilon P$, so that it only remains
to be shown that $P\ge 0$. But this follows from Lemma A1(i--iii) and the
definition of $\delta$, $\delta=\inf_{W_1\in S_{\rho_1}} \left|
\frac{\langle{W_1}\rangle_{\rho_2}}{\langle{W_1}\rangle_{\rho_1}} \right|$
. We can easily know $P$ is not finer than $\rho_1$ or it is separable. $\Box$

The previous lemmas provides us with a way of determining when an
entangled state is finer than another one. With this result, we are now at
the position of fully characterizing optimal entangled state.

\textbf{ Theorem A1:} $\rho$ is optimal iff for all projectors on product vectors $P$ and
$\epsilon>0$, $\rho'=(1+\epsilon)\rho-\epsilon P$ is not a super witness (not a legitimate entangled state).

{\em Proof:} (If) According to Lemma A2, there is no entangled state which is
finer than $\rho$, and therefore $\rho$ is optimal. (Only if) If $\rho'$
is an entangled state, then according to Lemma A2 $\rho$ is not optimal. $\Box$

For optimal indecomposable entanglement witnesses, there exists the case that $W$ is optimal, but $W^\Gamma$ is not optimal \cite{Ha12}. For bound states(optimal super indecomposable witnesses), there must exist the case that $\rho$ is optimal, but $\rho^\Gamma$ is not optimal.

For optimal decomposable entanglement witnesses and optimal free entangled states, we have the following result.

\textbf{Theorem A2.} If $W$ is an optimal entanglement witness and $Q=W^\Gamma$ is an entangled state, $Q$ is optimal.

\end{document}